\documentclass[prl,showpacs,twocolumn,superscriptaddress]{revtex4-1}

\usepackage{amsfonts,amsmath,amssymb,graphicx}
\usepackage[usenames,dvipsnames]{xcolor}

\usepackage{pdfpages}

\newcommand{\be}{\begin{eqnarray}}
\newcommand{\ee}{\end{eqnarray}}

\newcommand{\bra}[1]{\langle{#1}|}
\newcommand{\ket}[1]{|{#1}\rangle}

\newcommand{\op}[1]{\hat{#1}}

\newcommand{\szo}{\hat{\sigma}_{\mathrm{z}1}}

\newcommand{\szt}{\hat{\sigma}_{\mathrm{z}2}}
\newcommand{\Aa}{\hat{a}^\dag\hat{a}}
\newcommand{\ha}{\hat{a}}
\newcommand{\hA}{\hat{a}^\dag}
\newcommand{\wc}{\omega_\mathrm{r}}
\newcommand{\wa}{\omega_\mathrm{a}}
\newcommand{\wdr}{\omega_\mathrm{d}}
\newcommand{\dr}{\delta_\mathrm{r}}
\newcommand{\hc}{\mathrm{h.c.}}

\begin{document}

 \title{High-fidelity resonator-induced phase gate with single-mode squeezing}

\author{Shruti Puri}
\affiliation{D\'{e}partment de Physique, Universit\'{e} de Sherbrooke, Sherbrooke, Qu\'{e}bec, Canada J1K 2R1}
\author{Alexandre Blais}
\affiliation{D\'{e}partment de Physique, Universit\'{e} de Sherbrooke, Sherbrooke, Qu\'{e}bec, Canada J1K 2R1}
\affiliation{Canadian Institute for Advanced Research, Toronto, Canada}%

\date{\today}

\begin{abstract}
We propose to increase the fidelity of two-qubit resonator-induced phase gates in circuit QED by the use of narrowband single-mode squeezed drive. We show that there exists an optimal squeezing angle and strength that erases qubit `which-path' information leaking out of the cavity and thereby minimizes qubit dephasing during these gates. Our analytical results for the gate fidelity are in excellent agreement with numerical simulations of a cascaded master equation that takes into account the dynamics of the source of squeezed radiation. With realistic parameters, we find that it is possible to realize a controlled-phase gate with a gate time of 200 ns and average infidelity of $10^{-5}$. 
\end{abstract}


\maketitle

Taking advantage of pulse shaping techniques~\cite{motzoi:2009a} and increasing coherence times~\cite{paik:2011a}, single-qubit gate fidelity exceeding 99\% has been demonstrated with superconducting qubits~\cite{barends,chow:2012a}. Similar fidelities have been reported for two-qubit gates based on frequency tunable qubits~\cite{barends}. Tuning the qubit transition frequency is, however, sometimes undesirable or difficult~\cite{paik:2011a} and, for this reason, fixed-frequency two-qubit gates are being actively developed~\cite{rigetti:2005a,blais:2007a,majer:2007a,rigetti:2010a,chow:2011a,chow:2013a,cross:2015a}. Unfortunately, the fidelity of these all-microwave gates~\cite{corcoles:2015a} is still below that required for fault-tolerant quantum computation~\cite{raussendorf:2007a}.

\begin{figure}
\centering\includegraphics[width=0.85\columnwidth]{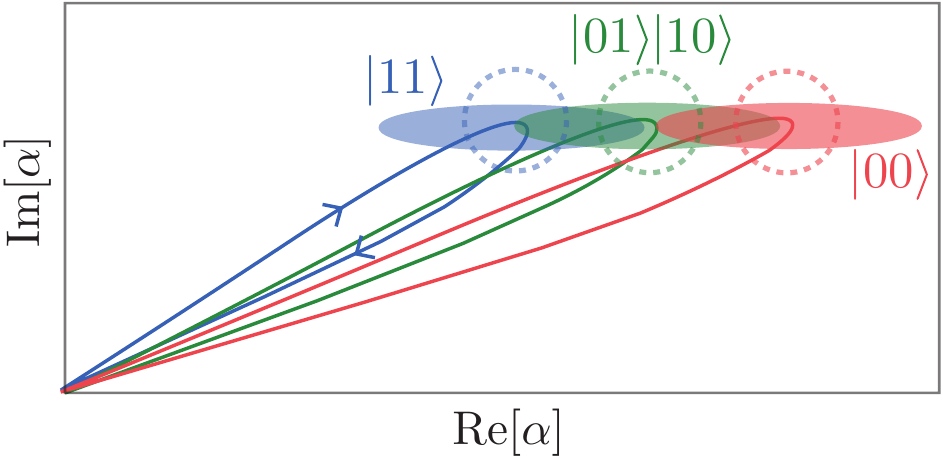}
\caption{(color online) Qubit-state-dependent evolution of the outgoing resonator field in phase space (solid lines) in the rotating frame of the drive. Quantum fluctuations corresponding to a coherent (dashed circles) and squeezed drive (filled ellipses) are represented at an extremum of the paths. By adjusting the squeezing angle, quantum fluctuations can help in erasing which-path information and thereby reduce qubit dephasing during the joint qubit-field evolution.}
\label{figPhaseCartoon}
\end{figure}

A promising all-microwave gate is the resonator-induced phase gate~\cite{cross:2015a}. This multi-qubit logical operation is based on the dispersive regime of circuit QED where the qubits are far detuned from a cavity mode~\cite{haroche:2006a,blais:2004a}. As schematically illustrated in Fig.~\ref{figPhaseCartoon}, adiabatically turning on and off an off-resonant drive, the cavity state evolves from its initial vacuum state by following a qubit-state-dependent closed loop in phase space. After this joint qubit-cavity evolution, the cavity returns to vacuum state and the qubits are left unentangled from the cavity but with an acquired a non-trivial phase. By adjusting the drive amplitude, frequency, and duration, an entangling phase gate can be realized~\cite{cross:2015a}. This is analogous to the geometric phase gate already demonstrated with ion-trap qubits~\cite{leibfried:2003a} and theoretically studied in the context of circuit QED, quantum dots in a cavity, and trapped ions~\cite{blais:2007a,puri,wang,garcia-ripoll:2003a}.

In practice, the gate fidelity is limited by residual qubit-cavity entanglement and by photon loss. Indeed, during the adiabatic pulse, photons entangled with the qubit leave the cavity carrying `which-path' information about the two-qubit state, in turn causing dephasing. This can be partially avoided by driving the cavity many linewidths from its resonance frequency. In this situation, the cavity is only virtually populated and the qubit-photon entanglement is small~\cite{blais:2007a,puri,wang,garcia-ripoll:2003a}. Unfortunately, this also leads to longer gate times, a problem that can be partially mitigated by using pulse shaping techniques~\cite{cross:2015a}.

Here we propose to use single-mode squeezing to address the challenge of implementing resonator-induced phase gates with a gate error below the fault-tolerance threshold and with short gate times. We show that an optimal, and experimentally realistic, choice of squeezing power and angle can dramatically improve the gate fidelity. The intuition behind this improvement is schematically illustrated in Fig.~\ref{figPhaseCartoon}: enhancing fluctuations in the appropriate quadrature erases the which-path information while leaving the path area, and hence the accumulated phases, unchanged. This improvement in gate fidelity is the converse of the recent realization that single-mode squeezed light is generally detrimental to dispersive qubit measurement~\cite{barzanjeh:2014a,didier:2015a}. Using squeezing powers close to that already experimentally achieved with superconducting circuits~\cite{eichler}, we find average gate errors that are suppressed by an order of magnitude with respect to a coherent state input drive. In other words, we suggest to use quantum-bath engineering to protect the dynamics of a quantum system, going beyond the typical use of this approach, which is focussed on creating or stabilizing certain steady-states~\cite{murch2012cavity,shankar2013autonomously,leghtas2015confining}.


In the dispersive regime where the qubit-cavity frequency detuning $\Delta = \wa-\wc$ is large with respect to the coupling strength $g$, the system Hamiltonian in the presence of a cavity drive takes the form 
\begin{equation}\label{eq:Hamil}
\begin{split}
\op{H} 
&=
\wc \Aa+\frac{\omega_\textnormal{a}}{2}\szo
+\frac{\omega_\textnormal{a}}{2}\szt+\chi(\szo+\szt)\Aa
\\&
+ \epsilon(t) (\hA e^{-i\wdr t} + \hc).
\end{split}
\end{equation}
In this expression, $\wc$ and $\wa$ are respectively the cavity and qubit frequencies, $\chi={g^2}/{\Delta}$ the dispersive coupling strength, $\epsilon$ the drive amplitude, and $\wdr$ the drive frequency. This description is accurate for intra-cavity photon number $n\ll~ n_\mathrm{crit}={\Delta^2}/{4g^2}$~\cite{blais:2004a}. Although the resonator-induced phase gate is tolerant to large variations in qubit frequencies and coupling strengths~\cite{SM}, to simplify the discussion we assume the qubits to be identical. 

How the resonator-induced phase gate emerges from evolution under Eq.~\eqref{eq:Hamil} can be made clearer by performing the time-dependent polaron-like transformation $D(\op\alpha')=\exp(\op\alpha' \hA-{\op{\alpha}'^*} \ha)$ with $\op\alpha'(t)=\alpha(t)-(\chi/\delta_{\text{r}})(\szo+\szt)\alpha(t)$ on $\op H$~\cite{mahan:2000a,gambetta:2008a}. As shown in the Supplemental Material~\cite{SM}, this leads to the effective Hamiltonian
\be\label{eq:Heff}
\op{H}_\textnormal{eff} = \tfrac{1}{2}\left[\wa+2\chi \op{n}(t)\right](\szo+\szt)-\frac{2\chi^2|\alpha|^2}{\dr} \szo\szt,
\ee
where we have defined $\op{n}(t) = \Aa + |\alpha(t)|^2$, with the amplitude $\alpha(t)$ satisfying $\dot\alpha=-i\dr\alpha-i{\epsilon(t)}$ and the drive-cavity detuning $\dr = \wc-\wdr$. 
The last term of Eq.~\eqref{eq:Heff} represents the nonlinear, qubit-state-dependent phase induced by the driven cavity. To avoid qubit-field entanglement after the gate, the cavity drive is chosen such that the field starts and ends in its vacuum state. For simplicity, we consider the drive to have a Gaussian profile $\epsilon(t) = \epsilon_0e^{-{t^2}/{\tau^2}}$ for times $-t_{\text{g}}/2<t<t_{\text{g}}/2$ with $t_{\text{g}}=5\tau$. As illustrated in Fig.~2(a), for $\dr\gg 1/\tau$ the cavity field evolves adiabatically and $\alpha(t)$ follows a closed path in phase space. With the cavity being only virtually populated, $\alpha(t)$ returns to the origin after the pulse.  
The qubit-state dependent phase acquired during this evolution is determined by the area in phase-space enclosed by $\alpha(t)$ and is specified by the pulse amplitude $\epsilon_0$, duration $\tau$, and detuning $\delta_{\text{r}}$~\cite{ladd2011simple}. By appropriately choosing these parameters, the evolution under Eq.~\eqref{eq:Heff} can correspond to the two-qubit unitary $U_{zz}=\mathrm{Diag}(1,1,1,-1)$. 

Another advantage of working with a large detuning $\dr$ is that measurement-induced dephasing $\gamma_\phi$ of the qubits is small~\cite{blais:2004a,measDeph}. On the other hand, the strength of the qubit-qubit interaction goes down with $\dr$, which in turns leads to long gate times. As we now show, we solve the challenge of minimizing $\gamma_\phi$ and maintaining short gate times by using an input field that is a displaced squeezed field rather than a coherent state. The squeezed field is characterized by the squeeze parameter $r(\omega)$ and angle $\theta$, and, in practice, can be produced by a Josephson parametric amplifier (JPA)~\cite{castellanos-beltran:2008a,eichler}. The frequency dependence of the squeeze parameter reflects the finite bandwidth of the JPA around the drive frequency $\wdr$.

Measurement-induced dephasing is caused by photon number fluctuations and, following Refs.~\cite{blais:2004a,measDeph, carm}, can be expressed as
\be\label{eq:deph}
\gamma_\phi(t)=2\chi^2\int^t_0 \left\langle [\op{n}(t)-\bar{n}(t)][\op{n}(t')-\bar{n}(t'])\right\rangle dt'.
\ee
An approximate expression for this rate can be obtained in the limit of adiabatic evolution of the cavity where the fast dynamics can be neglected. In this situation, this rate is given by~\cite{SM}
\be\label{eq:ideal}
\begin{split}
& \gamma_{\phi}(t) 
\approx \frac{4\chi^2 \kappa}{\dr^2}\\
&\times\left[N(\wc) + \frac{|\alpha(t)|^2}{2}\left(e^{-2r(\wdr)}\cos^2\Phi+e^{2r(\wdr)}\sin^2\Phi\right)\right],
\end{split}
\ee
where $\kappa$ is the cavity decay rate and $\Phi=\theta-{\text{arg}}[\alpha(t)]$ is the angle of squeezing relative to the drive.
We have also introduced $N(\wc)= \sinh^2 r(\wc)$, the thermal photon population associated with the squeezed input field. Crucially, this quantity is evaluated at the cavity frequency. This contribution to $\gamma_\phi$ can be made negligible by working at a detuning $\dr$ that is larger than the typically small bandwidth of current JPAs~\cite{castellanos-beltran:2008a,eichler}. Moreover, in the absence of squeezing ($r=0$), the second term of Eq.~\eqref{eq:ideal} is the usual expression for measurement-induced dephasing in a coherent field~\cite{measDeph}. While $N(\wc)$ in the first term is evaluated at the cavity frequency, the squeeze parameter in the second term is rather evaluated at the drive frequency $\wdr$. For $\dr\gg (1/\tau,\,\kappa)$, the cavity field closely follows $\alpha(t)\sim\epsilon(t)/\dr$ such that $\Phi\sim\theta$ at all times. As a result, choosing the squeeze angle $\theta = 0$ leads to an exponential reduction with increasing $r(\wdr)$ of the dephasing rate in the adiabatic limit. This confirms the intuition presented in Fig.~\ref{figPhaseCartoon} that increasing the quantum fluctuations in the appropriate quadrature with respect to the field displacement leads to a reduction of qubit dephasing.  Minimizing photon shot-noise by number-squeezed radiation was also studied in the context of cavity spin squeezing in Ref.~\cite{css}.

\begin{figure}
\includegraphics[width=0.9\columnwidth]{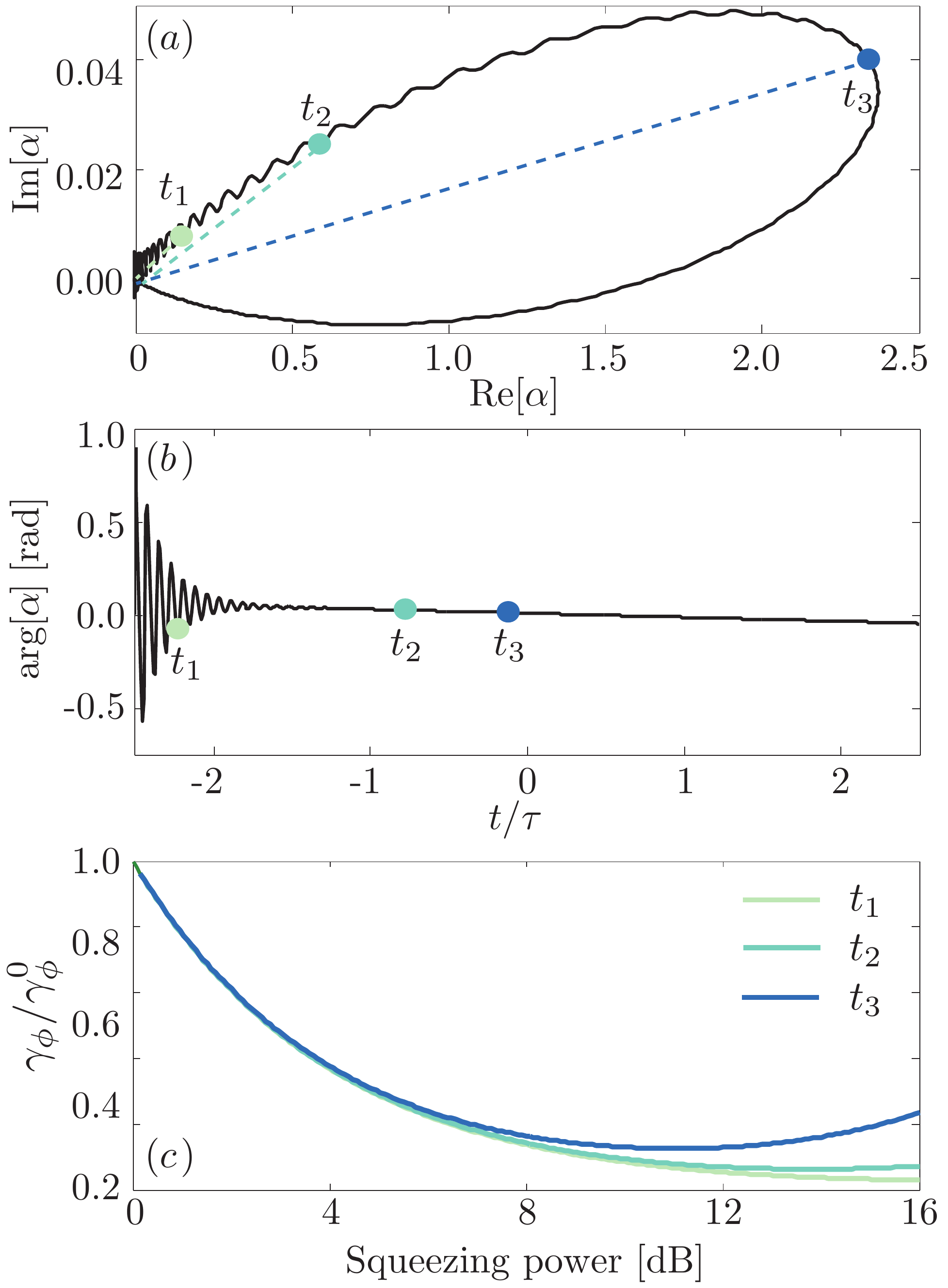}
\caption{(color online) (a) Evolution in phase space of the cavity field $\alpha(t)$ with $\dr/2\pi=320$ MHz, $\kappa/{2\pi}=10$ MHz, $\tau=40$ ns, and $t_\mathrm{g}=200$ ns. The colored dots represent the field at three different times and the corresponding lines represent the angles along which the quantum fluctuations should be reduced to minimize dephasing. (b) Time evolution of $\mathrm{arg}[\alpha]$. (c) Normalized dephasing rate evaluated at the three indicated times vs squeezing power for a fixed squeezing angle $\theta=0$ and $r(\wc)=0$. The normalization $\gamma_\phi^0$ corresponds to the situation without squeezing, i.e., $r=0$.}
\label{fig:angle}
\end{figure}

While the above argument suggests an exponential decrease of the dephasing rate for arbitrarily large squeezing powers, in practice there exists an optimal $r(\wdr)$. In order to understand this, it is useful to consider again the  evolution of the cavity field in phase space. As illustrated in Fig.~\ref{fig:angle}(a) and (b), for large detunings ${\text{arg}}[\alpha(t)]$ is small at all times, and choosing a constant $\theta\sim 0$ minimizes the dephasing rate. However, for large $r(\wdr)$, the anti-squeezed quadrature enhances dephasing at short times where ${\text{arg}}[\alpha(t)]$ fluctuates widely. This leads to an overall enhancement of $\gamma_\phi$. Figure~\ref{fig:angle}(c) shows the dependence of the normalized instantaneous dephasing rate on $r(\wdr)$ with $\theta=0$ at the three times indicated by the dots on Fig.~\ref{fig:angle}(a). The existence of an optimal squeezing power is clearly apparent.
The finite bandwidth of the input squeezed state is another reason for the existence of such an optimal point. Indeed, for a fixed squeezing bandwidth $\Gamma$, an increase in the squeezing power at $\wdr$ will also lead to an increase in thermal photons at $\wc$ with $N(\wc)=N(\wdr)/[(\omega-\wdr)^2+\Gamma^2]$~\cite{walls2007quantum}. This contributes to qubit dephasing via the first term of Eq.~\eqref{eq:ideal}. 
\begin{figure}
\includegraphics[width=0.9\columnwidth]{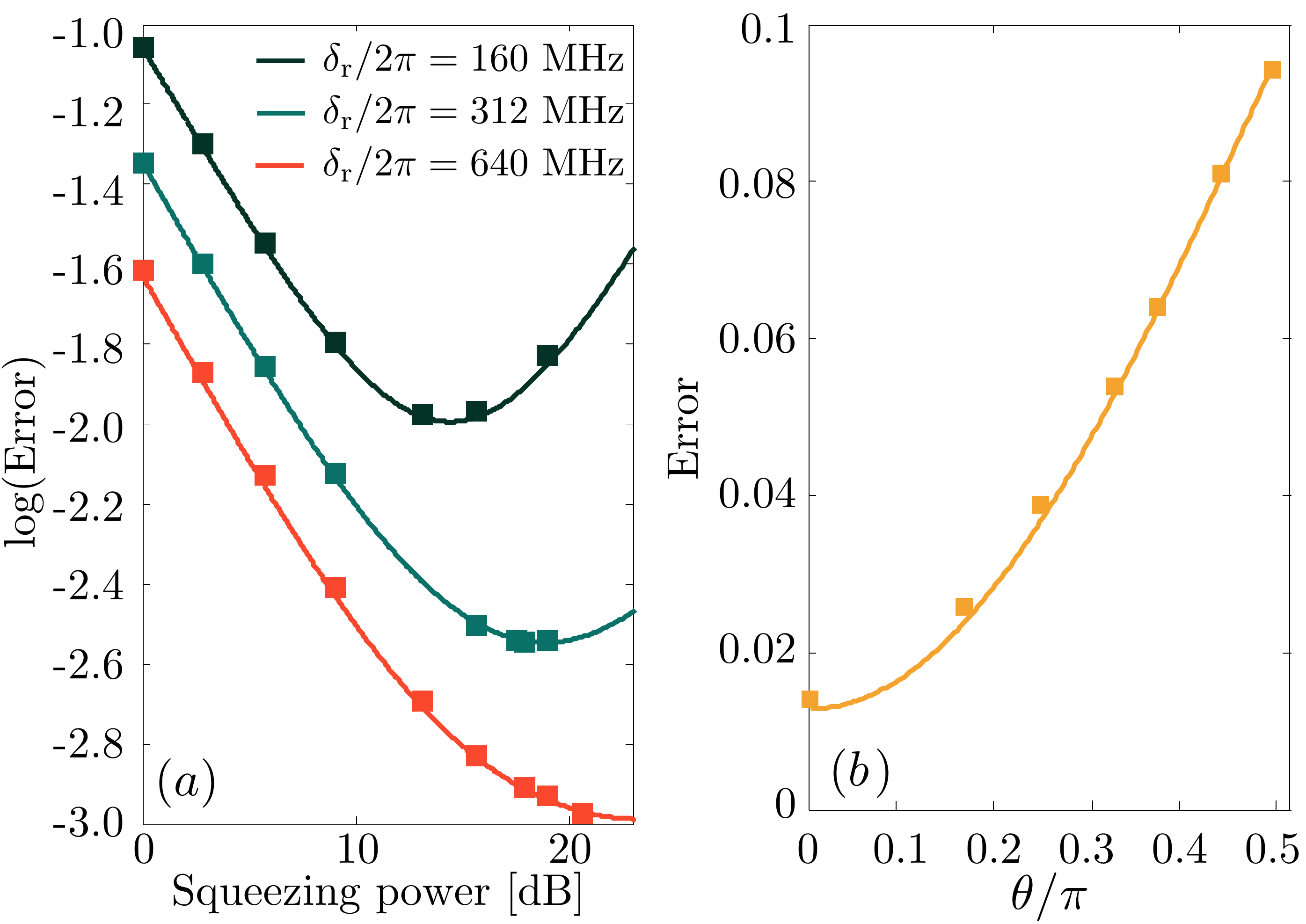}
\caption{(color online) (a) Analytical (solid lines) and numerical (symbols) gate error rate as a function of squeezing power for $\theta=0$ and three detunings ${\dr}/{2\pi}=160$ MHz (dark blue), $312$ MHz (blue) and $640$ MHz (red). The corresponding maximum drive amplitudes are $\epsilon_0/2\pi=278.5$ MHz, 795.8 MHz, and 2.31 GHz. The gate time is $t_\mathrm{g}=200$ ns, cavity decay ${\kappa}/{2\pi}=10$ MHz, qubit-cavity coupling ${g}/{2\pi}=160$ MHz, and detuning ${\Delta}/{2\pi}=3.2$ GHz corresponding to a dispersive coupling of ${\chi}/{2\pi}=8$ MHz and $n_\textnormal{crit.}=100$. (b) Analytical (solid lines) and numerical (symbols) gate error rate as a function of squeezing angle $\theta$ for a fixed squeezing power of 5.7 dB and detuning ${\dr}/{2\pi}=320$ MHz.}
\label{fig:error}
\end{figure}

We now turn to a more quantitative description of the improvement of gate fidelity that can be obtained from using squeezing. For this, we first compute the gate error $E= 1- \bra{\psi_{T}} \mathcal{E}(\ket{\psi_0}\bra{\psi_0})\ket{\psi_{T}}$ for the pure initial state $\ket{\psi_0}=\frac{1}{2}(\ket{00}+\ket{01}+\ket{10}+\ket{11})$ at $t=-t_\mathrm{g}/2$. In this expression, $\ket{\psi_{T}}=U_{zz}\ket{\psi_0}$ is the desired target state and $\mathcal{E}\cdot$ is the quantum channel representing the system under evolution with the Hamiltonian of Eq.~\eqref{eq:Hamil} in addition to the dephasing in the presence of a displaced squeezed drive. Following the notation from Ref.~\cite{cross:2015a}, the action of the channel on the qubits' density matrix elements takes the form  $\mathcal{E}(\ket{ij}\bra{kl})=e^{i\mu_{ij,kl}-\gamma_{ij,kl}}$, with $\{{i,j,k,l}\} \in \{0,1\}$, and where $\mu_{ij,kl}$ are qubit-state dependent phases and $\gamma_{ij,kl}$ represents non-unitary evolution due to the dephasing rate $\gamma_\phi$. In addition to two-qubit phases, evolution under the Hamiltonian of Eq.~\eqref{eq:Hamil} leads to single-qubit $z$ rotations. Since these rotations can be eliminated by an echo sequence, these are not considered in the error estimation~\cite{cross:2015a}. 

A prescription to evaluate $\mu_{ij,kl}$ and $\gamma_{ij,kl}$, and therefore the error $E$, can be found in~\cite{SM}. This corresponds to the full lines in Fig.~\ref{fig:error}(a) that show the error as a function of squeezing strength for different detunings $\dr$ and fixed $\theta = 0$. The symbols in this figure are obtained by numerical integration of the cascaded master equation for the system described by the Hamiltonian Eq.~\eqref{eq:Hamil} and driven by a degenerate parametric amplifier acting as a source of squeezed radiation~\cite{carmichael:1993b,SM}. The numerical simulations of the cascaded master equation are carried out with an open source computational package~\cite{johansson2012qutip,johansson2013qutip}. We have fixed $\tau=40$ ns which corresponds to a gate time of $t_\mathrm{g}=200$~ns. The drive parameters are chosen such that the cavity is empty at $t=t_\mathrm{g}/2$ and a maximum of 6 photons are excited in the cavity, which corresponds to $n/n_\mathrm{crit}=0.06$. The squeezing spectrum is centred at the drive frequency. Its linewidth $\Gamma/2\pi=32$ MHz is chosen to ensure that $\dr\gg\Gamma$, which minimizes the thermal photon population at the cavity frequency. As expected from the discussion above, the error goes down with increasing detuning. More importantly, the error is reduced in the presence of a squeezing input field up to a optimal power, beyond which measurement-induced dephasing again contributes to the gate error. As the detuning decreases, there is greater variation in $\mathrm{arg}[\alpha]$, which results in a reduction of the optimal squeezing power. Figure~\ref{fig:error}(b) shows the error as a function of the squeezing angle and confirms that the optimal choice is $\theta=0$

Table~\ref{tab:fidelity} presents the average gate fidelity obtained from numerical simulations with ($F^\textnormal{Sqz.}_\textnormal{avg}$) and without ($F^0_\textnormal{avg}$) squeezing for different detunings $\dr$ and cavity linewidth $\kappa$, but fixed gate time $t_\mathrm{g}=200$~ns~\cite{neil2, SM}. The parameters in the table are again chosen to limit the maximum number of photons in the cavity to $\sim 6$. It is important to emphasize the wide range of values chosen for $\kappa$ in this table. Note that in all cases the present approach leads to an increase in gate fidelity. Moreover, with squeezing powers close to what has already been realized experimentally~\cite{eichler}, an order of magnitude improvement can be obtained. Working with large detunings, an infidelity of $\sim 3.5\times 10^{-5}$, for example, can be obtained in a cavity with $\kappa/2\pi=50$~KHz. Notably, this is two orders of magnitude bellow the fault-tolerance threshold of the surface code~\cite{surfcode}. We also note that this approach can be combined with the improvement achieved by pulse shaping~\cite{cross:2015a}. Finally, we note that our scheme is robust to impure squeezing at the input. Indeed, because the infidelity depends only on the squeezed quadrature of the input drive, a thermal squeezed drive, or losses before the cavity, are simply equivalent to a reduction in the squeezing strength~\cite{SM}.

In summary, we have described a protocol to improve the fidelity of a two-qubit resonator-induced phase gate by over an order of magnitude. This improvement is based on which-path information erasure by using single-mode squeezing. The optimal squeezing strengths are close to what can already be achieved experimentally with superconducting quantum circuits. This scheme, based on tailoring the reservoir to dynamically protect a system during a logical operation, broadens the scope of quantum-bath engineering.

{\it Acknowledgements} - We thank Nicolas Didier and Arne~L.~Grimsmo for useful discussions. This work was supported by the Army Research Office under Grant W911NF-14-1-0078, INTRIQ and NSERC.

\begin{table}[t]
 \begin{tabular}{|c|c|c|c|c|c|c|c|}
 \hline
 $\frac{\dr}{2\pi}$&$\frac{\kappa}{2\pi}$ &$\chi$ &   $\epsilon_0$&$F^0_\textnormal{avg}$ & $F^\textnormal{Sqz.}_\textnormal{avg}$& Squeezing\\   
 (MHz)& (MHz)& (MHz) & (GHz) & $\%$& $\%$& Power (dB)\\ \hline  
320& 10  &8   &  0.796 & 98.16 & 99.89  & 16  \\ \hline
640& 10   & 8   & 2.31  & 98.96  & 99.95   &19 \\ \hline
111.4 & 0.05  & 4.5   & 0.294  & 99.96  & 99.9965   &15.7 \\ \hline
\end{tabular}
\caption{Average gate fidelity with and without squeezing ($F^\textnormal{Sqz.}_\textnormal{avg}$ and $F^0_\textnormal{avg}$, respectively). The gate $U_{zz}$ is implemented in time $t_{\text{g}}=200$ ns. 
}
\label{tab:fidelity}
\end{table}

\bibliography{twoqubit.bbl}{}
\clearpage


\section*{Supplementary material}

# Supplemental material for "High-fidelity resonator-induced phase gate with single-mode squeezing"

Shruti Puri[1] and Alexandre Blais[1,2]
1. Départment de Physique, Université de Sherbrooke, Sherbrooke, Québec, Canada J1K 2R1 and
2. Canadian Institute for Advanced Research, Toronto, Canada

## EFFECTIVE HAMILTONIAN

In the rotating frame of the drive the Hamiltonian of the two qubits dispersively coupled to a driven resonator reads,

$$\hat{H} = \delta_r \hat{a}^\dagger \hat{a} + \frac{\delta_{q,1}}{2}\hat{\sigma}_{1z} + \frac{\delta_{q,2}}{2}\hat{\sigma}_{2z} + (\chi_1\hat{\sigma}_{1z} + \chi_2\hat{\sigma}_{2z})\hat{a}^\dagger\hat{a} + \epsilon(t)(\hat{a} + \hat{a}^\dagger) \tag{S1}$$

with $\delta_{qi} = \omega_{ai} - \omega_d$ and $\chi_i$ the dispersive shift for $i = 1, 2$. On this Hamiltonian we apply the time-dependent displacement transformation $\hat{\tilde{H}} = D^\dagger(\hat{\alpha}')\hat{H}D(\hat{\alpha}') - iD^\dagger(\hat{\alpha}')\dot{D}(\hat{\alpha}')$, where $D(\hat{\alpha}') = \exp(\hat{\alpha}'\hat{a}^\dagger - \hat{\alpha}'^*\hat{a})$ and $\hat{\alpha}' = \alpha - (\chi_1\hat{\sigma}_{1z} + \chi_2\hat{\sigma}_{1z})\alpha/\delta_r$. This transforms the resonator operator to $\hat{a} \to \hat{\alpha}' + \hat{a}$, where $\hat{\alpha}'$ is the qubit-state-dependent coherent amplitude of the resonator mode and the quantum fluctuations are contained in the operator $\hat{a}$. Under this transformation with $\dot{\hat{\alpha}}' = -i\delta_r\hat{\alpha}' - i(\chi_1\hat{\sigma}_{1z} + \chi_2\hat{\sigma}_{1z})\hat{\alpha}' - i\epsilon(t)$ the Hamiltonian reduces to

$$\hat{H}_{\rm eff} = \delta_r \hat{a}^\dagger \hat{a} + \frac{\delta_{q1}}{2}\hat{\sigma}_{1z} + \frac{\delta_{q2}}{2}\hat{\sigma}_{2z} + (\chi_1\hat{\sigma}_{1z} + \chi_2\hat{\sigma}_{2z})\langle\hat{n}\rangle - 2\frac{\chi_1\chi_2|\alpha|^2}{\delta_r}\hat{\sigma}_{1z}\hat{\sigma}_{2z}. \tag{S2}$$

To simplify the above expression, we have taken the expression for $\hat{\alpha}'$ to zeroth order in $\chi$, in other words $\dot{\alpha} \sim -i\delta_r\alpha - i\epsilon(t)$ is a c-number in Eq. (S2). Taking $\hat{n} = |\alpha|^2 + \hat{a}^\dagger\hat{a}$ and simplifying to $\chi_1 = \chi_2 = \chi$, we recover the Hamiltonian of Eq. (2) of the main paper. It is important to note that although thermal resonator population (i.e. $\langle\hat{a}^\dagger\hat{a}\rangle$) contributes to the phase acquired by the qubits due to the dispersive interaction, a coherent resonator field ($\alpha$) is however required to mediate a two-qubit interaction between the qubits. Single qubit rotations arising from the fourth term of the above equation can be eliminated by spin echo and by choosing the drive pulse parameters i.e., the detuning, duration and power, it is possible to implement a gate given by the unitary operation $U_{zz} = \text{Diag}(1, 1, 1, -1)$.

## THEORETICAL ESTIMATION OF DEPHASING RATE

As described in the main paper, the dephasing rate is given by

$$\gamma_\phi = 2\chi^2 \int_0^t \langle(\hat{n}(t) - \bar{n}(t))\langle(\hat{n}(t') - \bar{n}(t'))\rangle\rangle dt' \tag{S3}$$

where $\hat{n}(t) = a^\dagger(t)a(t)$ and $\bar{n}(t) = \langle a^\dagger(t)a(t)\rangle$. Under the displacement transformation $D(\alpha)$ given above

$$\begin{aligned}\langle\delta n(t)\delta n(t')\rangle = &\, \alpha^*(t)\alpha^*(t')\langle\hat{a}(t)\hat{a}(t')\rangle + \alpha(t)\alpha(t')\langle\hat{a}^\dagger(t)\hat{a}^\dagger(t')\rangle + \alpha(t)\alpha^*(t')\langle\hat{a}^\dagger(t)\hat{a}(t')\rangle + \alpha^*(t)\alpha(t')\langle\hat{a}(t)\hat{a}^\dagger(t')\rangle \\ &+ \alpha^*(t)\langle\hat{a}(t)\hat{a}^\dagger(t')\hat{a}(t')\rangle + \alpha(t)\langle\hat{a}^\dagger(t)\hat{a}^\dagger(t')\hat{a}(t')\rangle + \alpha^*(t')\langle\hat{a}^\dagger(t)\hat{a}(t)\hat{a}(t')\rangle + \alpha(t')\langle\hat{a}^\dagger(t)\hat{a}(t)\hat{a}^\dagger(t')\rangle \\ &+ \langle\hat{a}^\dagger(t)\hat{a}(t)\hat{a}^\dagger(t')\hat{a}(t')\rangle - \langle\hat{a}^\dagger(t)\hat{a}(t)\rangle\langle\hat{a}^\dagger(t')\hat{a}(t')\rangle,\end{aligned} \tag{S4}$$

where $\delta n(t) = \hat{n}(t) - \bar{n}(t)$. In this displaced and rotating frame, the Langevin equations read [S1]

$$\frac{d}{dt}a = -(i\delta_r + \frac{\kappa}{2})\hat{a} - i\hat{\mathcal{F}}_{\rm in}e^{i\omega_d t} \tag{S5}$$

$$\frac{d\alpha}{dt} = -(i\delta_r + \frac{\kappa}{2})\alpha - i\epsilon(t), \tag{S6}$$

where $\hat{\mathcal{F}}_{\rm in}$ corresponds to the input quantum noise coupled to the resonator. In the presence of the squeezer, the correlation functions of this field are as usual: $\langle\hat{\mathcal{F}}_{\rm in}^\dagger(\omega)\hat{\mathcal{F}}_{\rm in}(\omega')\rangle = N(\omega)\kappa\delta(\omega - \omega')$, $\langle\hat{\mathcal{F}}_{\rm in}(\omega)\hat{\mathcal{F}}_{\rm in}^\dagger(\omega')\rangle = (N(\omega) + $

$1)\kappa\delta(\omega-\omega')$, $\langle\hat{\mathcal{F}}_{\text{in}}(\omega)\hat{\mathcal{F}}_{\text{in}}(\omega')\rangle = M(\omega)\kappa\delta(\omega+\omega')$, and $\langle\hat{\mathcal{F}}_{\text{in}}^\dagger(\omega)\hat{\mathcal{F}}_{\text{in}}^\dagger(\omega')\rangle = M^*(\omega)\kappa\delta(\omega+\omega')$, with $M = |M|e^{2i\theta}$ and $|M| \le \sqrt{N(N+1)}$ [S2].

Using these expressions, we can evaluate the field correlation functions entering the dephasing rate. For example,

$$\langle\hat{a}^\dagger(t)\hat{a}(t')\rangle = \int_{-t_g/2}^{t}\int_{-t_g/2}^{t'}\langle\hat{\mathcal{F}}_{\text{in}}^\dagger(s)\hat{\mathcal{F}}_{\text{in}}(s')\rangle e^{-i\omega_d(s-s')+i\delta_r(t-s)-\kappa(t-s)/2-i\delta_r(t'-s')-\kappa(t'-s')/2}\,\mathrm{d}s\,\mathrm{d}s'$$

$$= \int_{-t_g/2}^{t}\int_{-t_g/2}^{t'}\int_{-\infty}^{\infty} N(\omega)e^{i(\omega-\omega_d)(s-s')+i\delta_r(t-s)-\kappa(t-s)/2-i\delta_r(t'-s')-\kappa(t'-s')/2}\,\mathrm{d}\omega\,\mathrm{d}s\,\mathrm{d}s'$$

$$\sim \int_{-\infty}^{\infty} N(\omega)\frac{e^{i(\omega-\omega_d)(t-t')}}{(\omega-\omega_r)^2+\kappa^2/4}\,\mathrm{d}\omega \tag{S7}$$

The last expressions is obtained by eliminating fast oscillating and decaying terms. Similarly,

$$\langle\hat{a}(t)\hat{a}(t')\rangle \sim \int_{-\infty}^{\infty} M(\omega)\frac{e^{i(\omega-\omega_d)(t-t')}}{[i(\omega_r-\omega)+\kappa/2][i(\omega_r+\omega-2\omega_d)+\kappa/2]}\,\mathrm{d}\omega \tag{S8}$$

In order to evaluate the above expressions we assume that the squeezing is centred at frequency $\omega_d$ with bandwidth $\Gamma$ such that $N(\omega) = N(\omega_d)\Gamma^2/[(\omega-\omega_d)^2+\Gamma^2]$ and $M(\omega) = M(\omega_d)\Gamma^2/[(\omega-\omega_d)^2+\Gamma^2]$. If $\delta_r \gg \Gamma, \kappa$ then the product of Lorentzians in Eq. (S7) and (S8) can be written as a sum of Lorentzians and the two equations reduce to:

$$\langle\hat{a}^\dagger(t)\hat{a}(t')\rangle \sim N(\omega_d)\frac{\kappa\Gamma}{2(\delta_r^2+\kappa^2/4)}e^{-\Gamma(t-t')}, \tag{S9}$$

$$\langle\hat{a}(t)\hat{a}(t')\rangle \sim M(\omega_d)\frac{\kappa\Gamma}{2(i\delta_r+\kappa/2)^2}e^{-\Gamma(t-t')}. \tag{S10}$$

Following the same procedure, we find

$$\langle\hat{a}(t)\hat{a}^\dagger(t')\rangle \sim N(\omega_d)\frac{\kappa\Gamma}{2(\delta_r^2+\kappa^2/4)}e^{-\Gamma(t-t')} + e^{-i\delta_r(t-t')-\kappa(t-t')/2} \tag{S11}$$

$$\langle\hat{a}^\dagger(t)\hat{a}^\dagger(t')\rangle \sim M(\omega_d)^*\frac{\kappa\Gamma}{2(-i\delta_r+\kappa/2)^2}e^{-\Gamma(t-t')}. \tag{S12}$$

Assuming the output field of the squeezer to be Gaussian, $\langle\hat{a}^\dagger(t)\hat{a}^\dagger(t')\hat{a}(t')\rangle = \langle\hat{a}(t)\hat{a}^\dagger(t')\hat{a}(t')\rangle = \langle\hat{a}^\dagger(t)\hat{a}(t)\hat{a}^\dagger(t')\rangle = \langle\hat{a}^\dagger(t)\hat{a}(t)\hat{a}(t')\rangle = 0$ and using Wick's theorem the remaining terms in Eq. (S4) can be evaluate to be

$$\langle\hat{a}^\dagger(t)\hat{a}(t)\hat{a}^\dagger(t')\hat{a}(t')\rangle - \langle\hat{a}^\dagger(t)\hat{a}(t)\rangle\langle\hat{a}^\dagger(t')\hat{a}(t')\rangle = \langle\hat{a}^\dagger(t)\hat{a}(t')\rangle\langle\hat{a}(t)\hat{a}^\dagger(t')\rangle + \langle\hat{a}(t)\hat{a}(t')\rangle\langle\hat{a}^\dagger(t)\hat{a}^\dagger(t')\rangle \tag{S13}$$

The above expression can be evaluated using Eq. (S9)-(S12) and since we assume $\delta_r \gg \Gamma \gg \kappa$ it simplifies to,

$$\langle\hat{a}^\dagger(t)\hat{a}(t)\hat{a}^\dagger(t')\hat{a}(t')\rangle - \langle\hat{a}^\dagger(t)\hat{a}(t)\rangle\langle\hat{a}^\dagger(t')\hat{a}(t')\rangle \sim \frac{N(\omega_d)\Gamma^2\kappa}{\delta_r^4} \sim \frac{N(\omega_r)\kappa}{\delta_r^2} \tag{S14}$$

The expressions for the correlation functions, Eq. (S9)-(S14) can now be substituted in Eq. (S4) to obtain an explicit expression for the dephasing rate. A compact approximate expression for this rate can be obtained under an ideal adiabatic evolution of the resonator field i.e., when the duration of the drive pulse $\tau \gg 1/\delta_r$. Under this assumption, the field amplitude evolves slowly compared to other relevant energy scales in the system, so that $\alpha(t) \sim \alpha(t')$. In this situation we find

$$\gamma_\phi(t) \sim \left(\frac{2\chi^2\kappa|\alpha(t)|^2}{\delta_r^2+\kappa^2/4}\right)\left[e^{-2r(\omega_d)}\cos^2\Phi + e^{2r(\omega_d)}\sin^2\Phi\right] + \frac{4\chi^2\kappa}{\delta_r^2}\sinh^2 r(\omega_r) \tag{S15}$$

with $\Phi = \theta - \arg[\alpha(t)]$, $N(\omega_d) = \sinh^2 r(\omega_d)$ and $M(\omega_d) = \sinh r(\omega_d)\cosh r(\omega_d)$. Under ideal adiabatic evolution of the cavity field amplitude with $\delta_r \gg \kappa$, $\arg[\alpha(t)] \sim 0$ during the whole gate operation. Thus a squeezed input field with a constant squeezing angle $\theta = 0$ leads to an exponential decrease of the dephasing rate.

Note that the above expressions is derived for an ideal squeezed state satisfying $M(\omega_d) = \sqrt{N(\omega_d)[N(\omega_d)+1]}$. However this need not be the case and the above equation can be generalized to a thermal squeezed drive as,

$$\gamma_\phi(t) = \left(\frac{2\chi^2\kappa|\alpha(t)|^2}{\delta_r^2+\kappa^2/4}\right)\left[2N_0(\omega_d) + e^{-2r(\omega_d)}\cos^2\Phi + e^{2r(\omega_d)}\sin^2\Phi\right] + \frac{4\chi^2\kappa}{\delta_r^2}\sinh^2 r(\omega_r) \tag{S16}$$

where $N(\omega_d) = N_0(\omega_d) + N_1(\omega_d)$, $N_1(\omega_d) = \sinh^2 r(\omega_d)$ and $M(\omega_d) = \sinh r(\omega_d)\cosh r(\omega_d)$. Thus, at $\Phi = 0$, $r(\omega_r) = 0$ the dephasing rate decreases exponentially with effective squeezing strength $r_{\text{eff}} = -\ln[\exp(-2r(\omega_d)) + 2N_0(\omega_d)]/2$. As long as the unsqueezed thermal population $N_0(\omega_d)$ is small (for $r_{\text{eff}} > 0$), an exponential decrease in $\gamma_\phi$ is obtained.

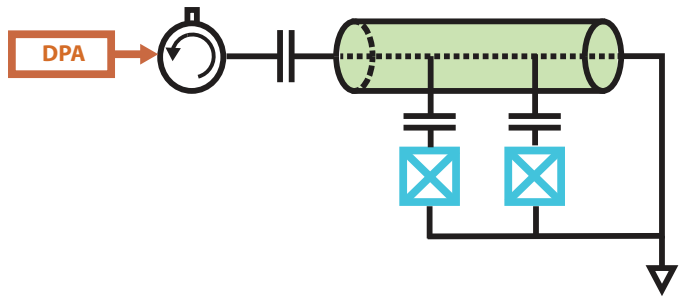

FIG. S1. Illustration of the setup proposed for implementing the two-qubit resonator-induced phase gate. A source of squeezing (DPA) drives a resonator that is dispersively coupled to two transmon qubits. The driven resonator induces a $\pi$-phase gate on the qubits.

## NUMERICAL ESTIMATION OF GATE FIDELITY

The symbols in Fig. 3 and the entries of Table 1 of the main paper are obtained by numerically solving the cascaded master equation for the system shown in Fig. S1. This corresponds to a resonator coupled to two qubits and driven by an ideal degenerate parametric amplifier (DPA). In frame rotating at $\omega_\mathrm{d}$, the cascaded master equation reads [S2]

$$\frac{d\hat{\rho}}{dt} = -i[\hat{H}, \hat{\rho}] + \mathcal{D}[\sqrt{\Gamma}\hat{b} + \sqrt{\kappa}\hat{a}]\rho, \tag{S17}$$

with

$$\hat{H} = \delta_\mathrm{r}\hat{a}^\dagger\hat{a} + i\epsilon_\mathrm{s}(t)(\hat{b}^{\dagger 2}e^{i\theta} - \hat{b}^2 e^{-i\theta}) + \epsilon(t)(\hat{a}+\hat{a}^\dagger) + i\frac{\sqrt{\Gamma\kappa}}{2}(\hat{b}^\dagger\hat{a} - \hat{a}^\dagger\hat{b}) + \frac{\delta_\mathrm{r}+\Delta}{2}(\hat{\sigma}_\mathrm{1z}+\hat{\sigma}_\mathrm{2z}) + \chi(\hat{\sigma}_\mathrm{1z}+\hat{\sigma}_\mathrm{2z})\hat{a}^\dagger\hat{a}, \tag{S18}$$

where $\hat{b}^{(\dagger)}$ are the annihilation (creation) operator for the DPA, $\Gamma$ linewidth of the DPA and $\theta$ determines the squeezing angle at the output of the DPA. The DPA is resonantly driven with a parametric drive $\epsilon_\mathrm{s}(t)$ of equal frequency to the coherent drive $\epsilon(t)$.

As discussed in the main text, for simplicity we use Gaussian pulse shapes for both the drives $\epsilon(t) = \epsilon_0(t)\exp(-t^2/\tau^2)$ and $\epsilon_\mathrm{s}(t) = \epsilon_{0,\mathrm{s}}(t)\exp(-(t-t_0)^2/\tau^2)$. It takes finite time for squeezing to develop in the DPA. As a result the parametric drive is displaced in time by a small amount ($t_0 \sim 8$ ns) to ensure that the peak of the coherent drive and that of the squeezing strength $r(t)$ coincide. At $t = -t_\mathrm{g}/2$, the DPA and the resonator are initialized to their vacuum state and the qubits to the separable state $|\psi_0\rangle = (|00\rangle + |01\rangle + |10\rangle + |11\rangle)/2$. The cascaded master equation is simulated using the computational package Qutip [S3],[S4]. Figure S2 shows the numerical estimate of the gate error for different squeezing powers when $\kappa \ll \chi$. From the figure we see that the qualitative dependence of the error on the squeezing power remains the same as when $\chi < \kappa$ (Fig. 3 in the main paper).

## ESTIMATION OF SQUEEZING POWER FROM DPA

Following Ref. [S2], the equation of motion for the DPA field reads

$$\frac{d\hat{b}}{dt} = -\frac{\Gamma}{2}\hat{b} + 2\epsilon_\mathrm{s}(t)\hat{b}^\dagger - \sqrt{\Gamma}\hat{b}_\mathrm{in}. \tag{S19}$$

Taking the Fourier transform of the above equation yields,

$$-i\omega\hat{b}(\omega) = 2\int \epsilon_\mathrm{s}(\omega-\omega')\hat{b}(\omega')d\omega' - \frac{\Gamma}{2}\hat{b}(\omega) - \sqrt{\Gamma}\hat{b}_\mathrm{in}(\omega). \tag{S20}$$

For a broad Gaussian pulse with duration $\tau > 1/\Gamma$, we can approximate the drive to be time-independent so that $\epsilon_\mathrm{s}(\omega-\omega') \sim \sqrt{\frac{2}{\pi}}\epsilon_{0,\mathrm{s}}\delta(\omega-\omega')$. In this case the correlations in the output field of a one-sided resonator are given

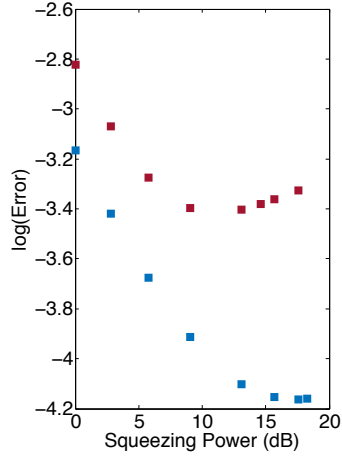

FIG. S2. Numerical result of gate error rate as a function of squeezing power for $\theta = 0$, $\delta_{\rm r}/2\pi = 56$ MHz, $\epsilon_0/2\pi = 104$ MHz (red) and 111 MHz, $\epsilon_0/2\pi = 295$ MHz (blue). The gate time is $t_{\rm g} = 200$ ns, $\tau = 40$ ns, $\kappa/2\pi = 50$ kHz, $\Gamma = 32$ MHz, $t_0 = 8$ ns and $\chi/2\pi = 4.5$ MHz

by [S2],

$$\langle \hat{b}_{\rm out}^\dagger(t)\hat{b}_{\rm out}(t')\rangle = \frac{\lambda^2 - \mu^2}{4}\left[\frac{e^{-\mu(t-t')}}{2\mu} - \frac{e^{-\lambda(t-t')}}{2\lambda}\right] \tag{S21}$$

$$\langle \hat{b}_{\rm out}^\dagger(t)\hat{b}_{\rm out}(t')\rangle = \frac{\lambda^2 - \mu^2}{4}\left[\frac{e^{-\mu(t-t')}}{2\mu} + \frac{e^{-\lambda(t-t')}}{2\lambda}\right], \tag{S22}$$

with $\lambda = \frac{\Gamma}{2} + 2\epsilon_{0,\rm s}$ and $\mu = \frac{\Gamma}{2} - 2\epsilon_{0,\rm s}$ so that,

$$N(\omega_{\rm d}) = \frac{(\lambda^2 - \mu^2)^2}{4\lambda^2\mu^2}, \qquad M(\omega_{\rm d}) = \frac{\lambda^4 - \mu^4}{4\lambda^2\mu^2}. \tag{S23}$$

Thus the squeezing power at the DPA resonator (or drive) frequency becomes

$$\text{Power} = -10\log(2N(\omega_{\rm d}) + 1 - 2|M(\omega_{\rm d})|) = 20r(\omega_{\rm d})\log(e). \tag{S24}$$

## ANALYTICAL GATE FIDELITY

As discussed in the main text, the full lines in Fig. 3 of the main text are obtained by analytically evaluating the dephasing rate. In this evaluation, we take advantage of the fact that the action of the gate on each element of the density matrix is $\mathcal{E}(|ij\rangle\langle kl|) = e^{i\mu_{ij,kl} - \gamma_{ij,kl}}$, $i,j,k,l = 0,1$, where $\mu_{ij,kl} = -2\chi^2(ij - kl)\int |\alpha(t)|^2 {\rm d}t/\delta_{\rm r}$ are the qubit-state dependent phases and $\gamma_{ij,kl}$ represents the non-unitary decay, $\gamma_{01,00} = \gamma_{10,00} = \gamma_{10,11} = \gamma_{01,11} = \int_{-t_{\rm g}/2}^{t_{\rm g}/2}\gamma_\phi(t){\rm d}t$, $\gamma_{01,10} = 0$ (for identical qubits) and $\gamma_{11,00} = 4\int_{-t_{\rm g}/2}^{t_{\rm g}/2}\gamma_\phi(t){\rm d}t$. $\gamma_{ij,kl}$ is obtained by substituting Eq. (S6) and (S9)-(S14) in Eq. (S4) and integrating numerically the resulting expression for $\gamma_\phi(t)$. We have neglected the single qubit rotations in the estimation of $\mu_{ij,kl}$ as they can be eliminated by an echo sequence.

As described in the main article, the gate fidelity is $F = {\rm Tr}[|\psi_{\rm 2T}\rangle\langle\psi_{\rm 2T}|\mathcal{E}(|\psi_0\rangle\langle\psi_0|)]$. Rather than evaluating the fidelity for a specific initial state, a more useful quantity is the entanglement fidelity [S5],

$$\mathcal{F}_{\rm Ent.} = \frac{1}{d^3}\sum_Q {\rm Tr}[Q\mathcal{U}^\dagger \mathcal{E}(Q)\mathcal{U}] \tag{S25}$$

where $Q$ is the set of $d \times d$ Pauli operators (with $d$ the dimension of the Hilbert space). In case of two qubits, this is the set of 16 operators $= \mathcal{I}, \sigma_{\rm x1}, \sigma_{\rm y1}, ...\sigma_{\rm x1}\sigma_{\rm x1},....$ $\mathcal{E}(Q)$ represents the action of the gate on each of these operators. We obtain the action of the gate on each element $|kl\rangle\langle mn|$, $\mathcal{E}(|k,l\rangle\langle m,n|) = e^{i\phi_{kl,mn} - \gamma_{kl,mn}}$ both numerically and

analytically following the above described approach. The average gate fidelity quoted in Table 1 of the main text is then obtained from [S6],

$$\mathcal{F}_{\text{avg}} = \frac{d\mathcal{F}_{\text{Ent.}} + 1}{d + 1} \tag{S26}$$

such that the average fidelity reads

$$\mathcal{F}_{\text{avg}} = \frac{1}{2} + \frac{1}{5}\left[\cos(\phi_{01,00})e^{-\gamma_{01,00}} - \cos(\phi_{11,01})e^{-\gamma_{11,01}}\right] - \frac{1}{10}e^{-\gamma_{11,00}}\cos\phi_{11,00}. \tag{S27}$$

---



\end{document}